# Metal Droplet Effects on the Composition of Ternary Nitrides

Mani Azadmand[1]*, Stefano Vichi[1], Sergio Bietti[1], Daniel Chrastina[2], Emiliano Bonera[1], Maurizio Acciarri[1], Alexey Fedorov[3], Shiro Tsukamoto[1], Richard Nötzel[4] and Stefano Sanguinetti[1]

1 QuCAT and Dipartimento di Scienza dei Materiali, Università di Milano-Bicocca, Milano (Italy)

2 L-NESS and Dipartimento di Fisica, Politecnico di Milano, Como (Italy)

3 L-NESS and IFN–CNR, Milano, (Italy)

4 QuCAT and Academy of Advanced Optoelectronics, South China Normal University, Guangzhou (China)

*Corresponding author: m.azadmand@campus.unimib.it

## Abstract

We investigate effects of metal droplets on the In incorporation in InGaN epilayers grown at low temperature (450 °C) by plasma assisted molecular beam epitaxy. We find a strong reduction of the In incorporation when the surface is covered by metal droplets. The such reduction increases with the droplet density and the droplet surface coverage. We explain this phenomenonology via a model that considers droplet effects on the incorporation of In and Ga adatoms into the crystal by taking into account the combined effects of the higher mobility of In, with respect to Ga, and to the vapor-liquid-solid growth that takes place under the droplet by direct impingement of nitrogen. The proposed model is general and can be extended to describe the incorporation of adatoms during the growth of the material class of ternary compounds when droplets are present on the surface.

## Introduction

InGaN is a direct band gap semiconductor with a wide tunability from 0.7 eV (InN) to 3.4 eV (GaN) depending on its actual stoichiometry[1]. In this way it is possible to cover almost the whole solar spectrum thus making InGaN the optimal material for solar applications[2]. InGaN characteristics are also desirable for many applications such as light sources, light detectors, photo electrochemical water splitting and electrochemical biosensors[3–6]. However the growth of InGaN films in the entire composition range is challenging for various reasons, mainly related to different thermal stabilities of In-N and Ga-N bonds, indium atom incorporation, and the miscibility gap in the InGaN alloy system[7,8]. The lattice mismatch of In-N and Ga-N leads to a miscibility gap which can cause fluctuations of the In content in the epilayer[8,9]. The calculated effects of strain in the InGaN bimodal and spinodal curves show that the miscibility problem remains significant for the case of InN mole fraction higher than 0.4[10]. InGaN phase separation has also been demonstrated experimentally, for both plasma assisted molecular beam epitaxy (PAMBE)[11] and metalorganic chemical vapor epitaxy (MOCVD) growth[12]. On the other hand, the different In-N and Ga-N bond energies are reflected in the different decomposition temperatures of InN (630°C) and GaN (850°C) [13]. Consequently, a reduction of In incorporation in the epilayer occurs not only due to the re-evaporation of physisorbed surface adatoms but also due to the thermal decomposition of In-N bonds. Hence, at usual InGaN growth temperatures, namely ~650 °C for MBE and ~800 °C for MOCVD, the incorporation of indium atoms is insufficient to achieve high indium concentrations[7,14–16].

Low temperature regime has been used to avoid InN decomposition and In desorption, and consequently to increase the In concentration in the crystal [17]. However, at low growth temperatures, the generation of metal droplets is critical since the excess metal atoms do not leave the surface[18], as has been reported in the case of GaN[19,20], unless the growth is carried out under N-rich conditions which will cause 3D growth and a rough surface[21,22].

Attempts to increase the In content of the crystal by raising the In flux results in formation of In droplets on the surface. Shimizu et al.[23] realized that for short growth times there is little to no In droplet generation (even at high In pressure). So they suggested that the growth of a thin GaN layer, when the In droplets are not generated or still small, can prevent the formation of In droplets and consequently increases the final In composition. Although the reason behind was not explained clearly, they reported a sizeable increase of the indium mole fraction to 29% with respect to the 12% achieved in a single layer grown under the same conditions[23].

During InGaN epilayer deposition performed by PAMBE at low growth temperatures and metal-rich conditions, the presence of droplets on the surface deeply affects the growth dynamics by depleting the metal adatoms on the surface[18]. Under these conditions, in addition to the expected metal incorporation at active sites on the surface, the VLS growth mode takes place under the droplets. Here we show that the presence of metal droplets on the surface, in addition to strong effects on the growth rate[18], also affects the incorporation of In and causes phase separation in the InGaN layer. We attribute this phenomenon to the combined effect of the different incorporation rates between In and Ga under the droplet by VLS and to the higher capture efficiency of In adatoms in between the droplets caused by the In larger diffusivity when compared to Ga[24,25].

## Experimental:

Si(111) wafers are used as the substrate in this study. The InGaN thin-films were grown using a MBE equipped with radio frequency (RF) plasma source. Oxide desorption was performed by thermal treatment at 850 °C in vacuum (background pressure < $5\times10^{-10}$ Torr) for 30 min. The oxide removal was confirmed by observing clean Si(111)-$1\times1\leftrightarrow7\times7$ surface reconstruction using in-situ reflection high-energy electron diffraction (RHEED) which is also used for the calibration of substrate temperature[26].

Due to the large lattice mismatch between Si and InGaN, prior to the InGaN growth, Si is exposed to active nitrogen flux for 5 min at 800°C, to obtain a thin $SiN_x$ layer which is known to improve the crystal quality of (In)GaN[3,27] layers. The RF source was operated with 0.9 sccm (standard cubic centimeter per minute) $N_2$ flux and RF power of 360 W. Subsequently the temperature of the sample was reduced to 450 °C. The growth of the InGaN thin-films were performed keeping the RF source parameters constant (N flux of $0.98\times10^{14}$ atoms $cm^{-2}$ $s^{-1}$) and total metal flux ($F_{Ga} + F_{In}$) varying between $0.98-2.94\times10^{14}$ atoms $cm^{-2}$ $s^{-1}$, keeping one to one ratio between In and Ga fluxes (see table 1). Fluxes have been calibrated by careful measurement of the thickness of In/GaN thin-films grown in full condensation regime. In case of metal fluxes, the calibration is based on the growth under N-rich conditions, and in case of N flux, the calibration is based on the growth in the intermediate growth regime (transition between N-rich and metal-rich). It is worth noting that that the typical calibration of the N flux under metal-rich conditions will be strongly affected by metal droplets as is discussed in our other paper[18].

The surface coverage by metal droplets was investigated by optical microscopy and scanning electron microscopy (SEM).

X-ray diffraction (XRD) was carried out using a PANalytical X'Pert PRO high-resolution diffractometer. The $K\alpha1$ radiation from the Cu anode ($\lambda$ = 0.15406 nm) was selected using a hybrid mirror and 2-bounce Ge monochromator. The sample was mounted on a high-precision goniometer with translational (*x*, *y* and *z*) and rotational (incidence angle ω, diffraction angle 2θ, sample rotation φ and sample tilt χ) degrees of freedom. A three-bounce Ge monochromator was placed in front of the detector as an analyser crystal, in order to obtain high precision in 2θ and to reject fluorescence from the sample. ω-2θ scans of the InGaN(0002) peak were obtained, using the Si(111) peak from the substrate as a reference. The In content of the $In_xGa_{1-x}N$ layer was estimated by linear interpolation between the c-plane lattice parameters of InN and GaN.

To investigate the local variation in the composition[28], line scan Raman spectroscopy was obtained using Micro Raman with an excitation lambda of 532 nm, excitation power of ≈ 2.5 mW, and spot diameter on the sample of 0.7 μm.

# Results and Discussion

Asymmetric reciprocal space maps by X-ray diffraction demonstrated that the InGaN layers are relaxed and maintain an epitaxial relationship with the substrate. Table 1 reports the growth conditions, surface area covered by metal droplets, and average In composition of each sample in the series, as deduced from X-ray diffraction measurement.

Samples M1, M2, and M3 are grown by supplying the same metal and nitrogen fluxes (III/V ratio $v$=1). Nevertheless, we observe a difference in the surface coverage by metal droplets of these three samples. Due to the low temperature conditions, even a small fluctuation in the growth parameters (temperature and metal and N fluxes) at $v$=1 is enough to change the growth toward N-rich or metal-rich conditions. We therefore interpret the difference in the droplet coverage of samples M1-M3 as due to the imbalances in the III/V ratio due to small fluctuations in the growth parameters.

Samples M6, M7, and M8, grown at the same conditions at $v$=3, show a change in the droplet coverage. This difference can be attributed to the different growth time, which increases the total dose of metal supplied to the samples.

The ω-2θ scan X-ray diffraction patterns of the samples with different surface coverage by metal droplets are shown in Fig. 1. As can be seen, there is a clear shift of the InGaN(0002) diffraction peak position to higher degrees, which corresponds to reduction of In incorporation from 53% (sample with no droplet at $F_{In}=F_{Ga}\approx 0.5\Phi$) down to 13±2% (sample with 12.7% surface coverage at $F_{In}=F_{Ga}=0.7\Phi$) and to 8±2% (sample with 25.2% surface coverage at $F_{In}=F_{Ga}=1.5\Phi$). The higher the metal flux, the lower becomes the In incorporation in the crystal. The common interpretation of this effect relates to the larger GaN formation enthalpy with respect to InN[29]. The deficiency of N with respect to the total metal flux, makes the In concentration $\zeta_{In}$ dependent on the difference between nitrogen and gallium fluxes: $\zeta_{In} = 1 - \frac{F_{Ga}}{\Phi}$. At $F_{Ga}>\Phi$, no In incorporation is expected. This linear dependence of $\zeta_{In}$ on $F_{Ga}$ is not observed in our samples. A strong decrease, from $\zeta_{In} = 0.53$ to 0.28, happens for Ga fluxes nominally equal to $F_{Ga}=0.5\Phi$, thus showing that the dependence of $\zeta_{In}$ on $F_{Ga}$ is much stronger than predicted.

Inspecting the grown surface, we observe that the appearance of droplets (Fig. 2) marks the onset of the reduction in In concentration. In addition to the expected dependence on the metal flux, we observe a more subtle effect. The In incorporation has a monotonous dependence with the surface coverage by metal droplets χ (Fig. 3), decreasing as the surface coverage increases. Such dependence shows two regimes: i) at low droplet coverage, and low excess metal flux, the $\zeta_{In}$ has a strong decrease with the increasing χ, going rapidly from 52% to ≈15% when droplets reach a coverage of χ=0.05. Additional increase in the droplet coverage reduces the total amount of indium in the crystal, but with a much slower dependence. At the droplet coverage of 38% the In concentration in the InGaN epilayer is reduced to 6%. Thus the presence of droplets on the surface and their actual surface coverage appear to have a relevant role in determining $\zeta_{In}$ in the bulk InGaN.

As recently reported[18], metal droplets have strong effects on the growth rate of InGaN epilayers, which experiences a continuous reduction as the overall metal flux impinging on the surface increases. This effect has been attributed to the reduction of the adatom density, and thus, in turn, of the growth rate, owing to the activation, in the presence of droplets, of the process of adatom attachment and incorporation into the droplets. This depletion channel is in competition with the N-driven metal adatom incorporation into the InGaN crystal. This leads, on one side, to a reduction of the metal adatom incorporation rate into the crystal in the regions not covered by the droplets. On the other side, the presence of liquid metal droplets on the surface should allow for the vapour liquid solid growth of InGaN material under the droplets themselves, via direct incorporation of nitrogen. In summary, according to Azadmand et al. [18], four processes are available to the metal adatom on the growing InGaN surface in the presence of droplets:

1) Desorption
2) Incorporation of metal adatom into the crystal by binding to a N active site on the droplet-free surface
3) Metal adatom attachment to a droplet
4) Incorporation via VLS at the droplet footprint.

In order to determine the effect of the combination of these phenomena on the actual In incorporation, we should resolve first the equations describing the kinetics of the In and Ga adatom densities $n_{Ga}$ and $n$ in the areas in between the droplets, thus taking into account for processes 1-3:

$$\frac{dn_{Ga}}{dt} = F_{Ga} - E_{Ga} - n_{Ga}\delta(\Phi) - n_{Ga}\sigma_D\rho D_{Ga} \qquad (1)$$

$$\frac{dn_{In}}{dt} = F_{In} - E_{In} - n_{In}\delta(\Phi) - n_{In}\sigma_D\rho D_{In} \qquad (2)$$

where $E_{Ga}$ and $E_{In}$ are the desorption fluxes of Ga and In, respectively, and $\delta(\Phi)$ the incorporation probability into the crystal on the droplet free surface, which depends on the flux $\Phi$. $\delta(\Phi)$ is proportional to the probability, for a metal (Ga or In) adatom, to find an active N site for binding. $\rho$ is the droplet density and $\sigma$ the droplet capture cross section. $D_{Ga}$ and $D_{In}$ are the Ga and In diffusivities, respectively. We cannot access experimentally the exact density of droplets acting during the growth. The droplet density that is measured after the growth, especially at high coverage, is affected by Ostwald ripening effects that take place during the cooling of the sample and it is therefore not related to the droplet density in the growth regime. We will therefore explicitly use the information $\rho$ only when $\chi \ll 1$. At low temperature in MBE conditions the metal desorption flux for both In and Ga is close to zero (complete condensation regime), so $E_{Ga}$ and $E_{In}$ can be neglected. In steady state conditions, and considering that in our conditions $F_{Ga} = F_{In}$, Eqs (1) and (2) can be reduced to the following

$$\frac{n_{In}}{n_{Ga}} = \frac{\sigma_D \rho D_{Ga} + \delta}{\sigma_D \rho D_{In} + \delta} \qquad (3)$$

As demonstrated in Ref.[18], the capture rate of the droplets increases with the droplet density. From the point of view of the In and Ga adatom densities, being $D_{In} \gg D_{Ga}$, this results in a strong imbalance of the density towards an excess of Ga atoms as the density of droplets on the growth surface increases. As droplet attachment exceeds the incorporation rate, the In and Ga adatom ratio is equal to:

$$\frac{n_{In}}{n_{Ga}} \sim \frac{D_{Ga}}{D_{In}} \qquad (4)$$

Because $D_{In} \gg D_{Ga}$ [24,25], from Eq. (4) we obtain that $n_{Ga} \gg n_{In}$. That is, due their larger diffusivity, In adatoms are more likely captured by the droplets, this way depleting the In adatom density on the surface. We expect, therefore, a strong reduction of the In content in the crystal grown in between the droplets as soon as the droplets are formed on the surface. Thus, this effect accounts for the drastic drop of In concentration at low droplet coverage to values around

$$\zeta_{in} \sim D_{Ga}/(D_{In} + D_{Ga}) \qquad (5)$$

The reduction of In incorporation due to the presence of metal droplets on the surface also has been observed by others[15,30] and related to the InN bond instability related to the high growth temperature. The low grow temperature used here rules out this as a possible interpretation of the experimental observation.

As a matter of fact, the In adatom depletion by preferential droplet attachment cannot account for the additional reduction, with a moderate dependence on the droplet coverage, that takes place at higher metal fluxes. On the other side, we expect that, as soon as droplets start to form on the substrate, a second growth channels starts on the surface, due to the VLS process that takes place under the droplets (process 4)[18].

As a matter of fact, when a droplet composed by of two metals, as we expect to be the droplet formed on the surface by the contemporaneous irradiation of In and Ga, is irradiated with a flux of a group V element, the VLS growth mode that takes place at the liquid-solid interface leads to the segregation, in the epitaxial layer, of the metal with the higher reactivity with the group V element of the two. In order to quantify this effect, Priante et al.[31] introduced the following expression to connect, in bi-metallic droplets, the compositions of the more reactive element (here Ga) in the liquid droplet ($u_{Ga}$) and in the solid ($\zeta_{Ga}$):

$$\zeta_{Ga}(u_{Ga}) = \frac{u_{Ga}\epsilon}{[1 + (\epsilon - 1)u_{Ga}]} \qquad (6)$$

where $\varepsilon$ (which depends only on temperature) is the ratio between the reaction quotients of the crystallization of the metal species (here Ga and In) with group V elements (here N) in the liquid: $Ga^l + N^l \rightarrow GaN^s$ and $In^l + N^l \rightarrow InN^s$. In case of Ga and In, being the enthalpy of formation of GaN and InN $\Delta H_{GaN} = 156.8 \pm 16\ kJ/mol$ and $\Delta H_{InN} = 28.6 \pm 9\ kJ/mol$, $\varepsilon$ 220. Thus we expect, due to the segregation effect induced by VLS, an increase of the Ga concentration $\zeta_{Ga}$ in the layer growing under the droplet with respect to the composition $\alpha$ expected from the Ga and In fluxes ratio impinging on the surface. As a rule of thumb, the excess of Ga, with respect to the steady state conditions, will be segregated at the interface, leaving the excess of In in the droplet. It is possible to have an estimate of the composition of the growing epilayer in VLS conditions by solving the equation[31]

$$du_{Ga}/dt = -G\,\zeta_{Ga}(u_{Ga}) + K \qquad (7)$$

where G is the incorporation rate and K is the Ga flux, both normalized to the droplet volume. In droplet growing conditions, that is whenever the flux of the each metal elements on the droplet is higher than the incorporation rate at the bottom of the droplet $(-G\zeta_{Ga}(u_{Ga}) + K \gg 0)$, the solution of (7), for $\varepsilon \gg 1$, is

$$u_{Ga} \gg \frac{K}{\varepsilon(G-K)} \qquad (8)$$

Combining Eqs. (6) and (8), the Ga concentration $\zeta$ in the InGaN layer growing by VLS under the droplet is $\zeta_{Ga} \gg \frac{K\varepsilon}{(G\varepsilon - K)}$ which, being $\varepsilon \gg 1$, reduces to

$$\zeta_{Ga} \gg K/G \qquad (9)$$

In droplet growing conditions, $K > \alpha G$ so that the concentration of Ga under the droplet is $\zeta_{Ga} \gg \alpha$. As a consequence, the VLS growth under the droplets strongly favors the Ga segregation at the droplet footprint, which can be considered, due to the high value of ε, close to pure GaN. This will induce a decrease with the droplet coverage of the average In content of the grown epilayer. This is what is observed. An oversimplified model, assuming $\zeta_{In} \sim 0$ under the droplets and a constant concentration, set by diffusivity ratio in between the droplets, leads to the following equation relating the average In composition in the epilayer with the droplet coverage:

$$\zeta_{In}(\chi) = (1 - \chi)\zeta_{In}^s \qquad (10)$$

Where $\zeta_{In}^s$ is the concentration of In in the droplet surroundings indicated by Eq. (5), thus in the conditions of zero droplet coverage but with the adatom depletion channel active. The blue line in Fig. 3 reports the expected behavior of $\zeta_{In}(\chi)$ assuming $\zeta_{In}^s = 0.15$. The slow reduction of the In concentration at high droplet coverage could be therefore attributed to the presence of a sizeable segregation of Ga at the droplet footprint. The $\zeta_{In}^s = 0.15$ corresponds to In diffusivity six times larger than that of Ga at the growth temperature (450 °C).

In order to experimentally check the results of Eq. (9), that is a strong increase of the Ga concentration below the droplet with respect to the surrounding material, we mapped the In concentration locally by micro-Raman spectroscopy. In samples grown in metal-rich condition, it is possible to find droplets' footprints on the surface left uncovered from metal by the effect of Ostwald ripening[32] which takes place during the cooling of the substrate. This allows for the investigation of the stoichiometry of the crystallized layer at the bottom of the droplet. We found a clear increase in the Ga concentration in the droplet footprint with respect to the surrounding areas (see Fig. 4). The Raman line-scan spectra around the footprint of a metal droplet shows a shift in the position of the $A_1$ (LO) peak to higher frequencies as the laser beam moves from the area surrounding the droplet to its footprint. The observed Raman shift corresponds to In concentration[28] in flat area $\zeta_{In} = 0.15$ and $\zeta_{In} = 0.07$ under the droplet (≈ 7%) thus qualitatively confirming the Eq.(9) predictions.

Confirmation of the phase separation and compositional gradients in the epilayer which may be caused by VLS growth under the droplet comes from XRD and Raman spectroscopy of the samples. In Fig. 5, The broadening of the ω–2θ scan XRD peaks of the InGaN samples with different χ are shown. It is a result of variation in lattice parameter due to different composition under the droplets with respect to the flat area between the droplets (as it has been explained before and confirmed by Raman line scan around and across the droplet footprint). This broadening is observed in the case of all InGaN samples with considerable χ values, and is different from the broadening observed in case of InGaN sample with indium concentration around 53% which is due to the immiscibility of InN and GaN. On the contrary, the samples with low value of χ (χ<0.05), thus only slightly affected by VLS effects, although displaying a strong reduction in the In content, show a single peak.

## Conclusions

The presence of droplets on the surface of InGaN ternary nitrides gives rise, in addition to the already observed reduction in the growth rate[18], to an overall reduction of the In percentage in the layer. At high droplet coverage clear fluctuation in the InGaN composition has been observed. These effects are related to the two distinct processes triggered by the presence of droplets: 1) adatom attachment to the droplet, with the consequent reduction of the free adatom density for the growth; 2) the VLS growth mode that takes place at the droplet-substrate interface.

The same two processes were at the basis of the droplet induced quenching of the growth rate. Their effect on the InGaN composition stems from the different magnitude of their effect on In and Ga adatoms. The higher mobility of In makes it more prone to be captured by droplet, thus depleting $n_{In}$ between the droplets. The much higher formation enthalpy of GaN, with respect to InN, favors the segregation of GaN at the bottom of the droplets, thus leaving unreacted indium in the droplet. The combination of the two effects, well describes the observed reduction of In incorporation with the metal flux and the surface coverage.

In addition to the reduction of the incorporation of In into the epilayer the presence of droplets induces a surface fluctuation in the In content, due to the VLS growth mode favoring the Ga segregation at the interface between the droplet and the substrate. These findings permit to understand, in terms of competition between VLS and normal growth modes, the observed limits in the film uniformity of two recently proposed growth modes of ternary nitrides based on the use of metal droplets, the Metal Modulated Epitaxy (MME)[33,34] and the droplet elimination by radical beam irradiation (DERI)[35,36], which were managed through a careful control of the adatom kinetics.[37,38] According to our findings, droplet surface coverage is the fundamental factor affecting uniformity in MME and DERI. In the presence of droplets segregation effects, related to the VLS growth mode, cannot be avoided. Still they can be minimized by a careful control of the droplet coverage $\chi$.

Table 1. Growth parameters and samples characteristics

| Samples name | In+Ga metal flux ($10^{14}$ atoms cm$^{-2}$ s$^{-1}$) | III/V ratio | Growth time (min) | Density of metal droplets ($10^6$ cm$^{-2}$) | Surface coverage by metal droplets (%) | In concentration (%) |
|---|---|---|---|---|---|---|
| M1 | 0.98 | 1 | 60 | 0 | 0 | 53 |
| M2 | 0.98 | 1 | 120 | 1.6 | 0.4 | 31 |
| M3 | 0.98 | 1 | 60 | 31 | 5.6 | 28 |
| M4 | 1.41 | 1.4 | 90 | 34 | 12.7 | 13 |
| M5 | 1.96 | 2 | 45 | 91 | 17.8 | 13 |
| M6 | 2.94 | 3 | 45 | 7 | 25.2 | 8 |
| M7 | 2.94 | 3 | 90 | 6.1 | 36.3 | 4 |
| M8 | 2.94 | 3 | 150 | 5.4 | 37.4 | 5 |

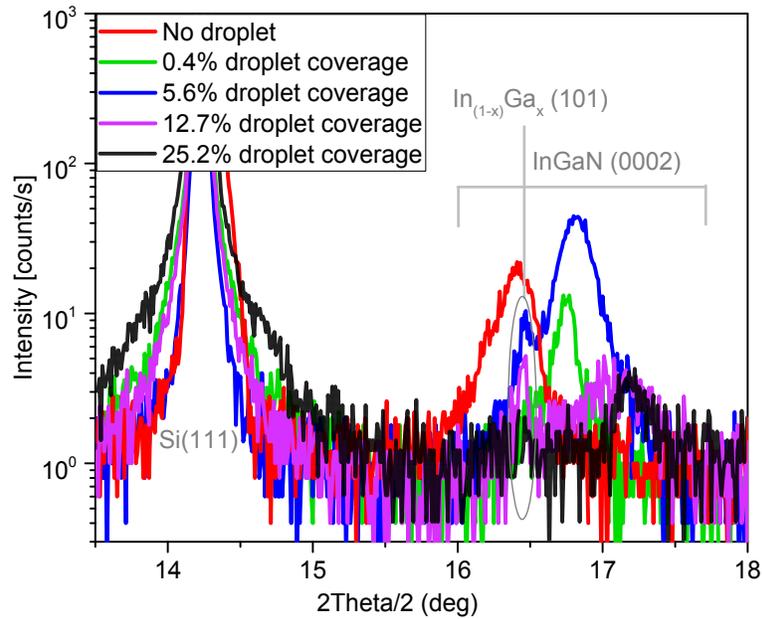

Fig.1. ω-2θ scan XRD of InGaN thin-films (grown on Si(111)) with different metal droplets surface coverage.

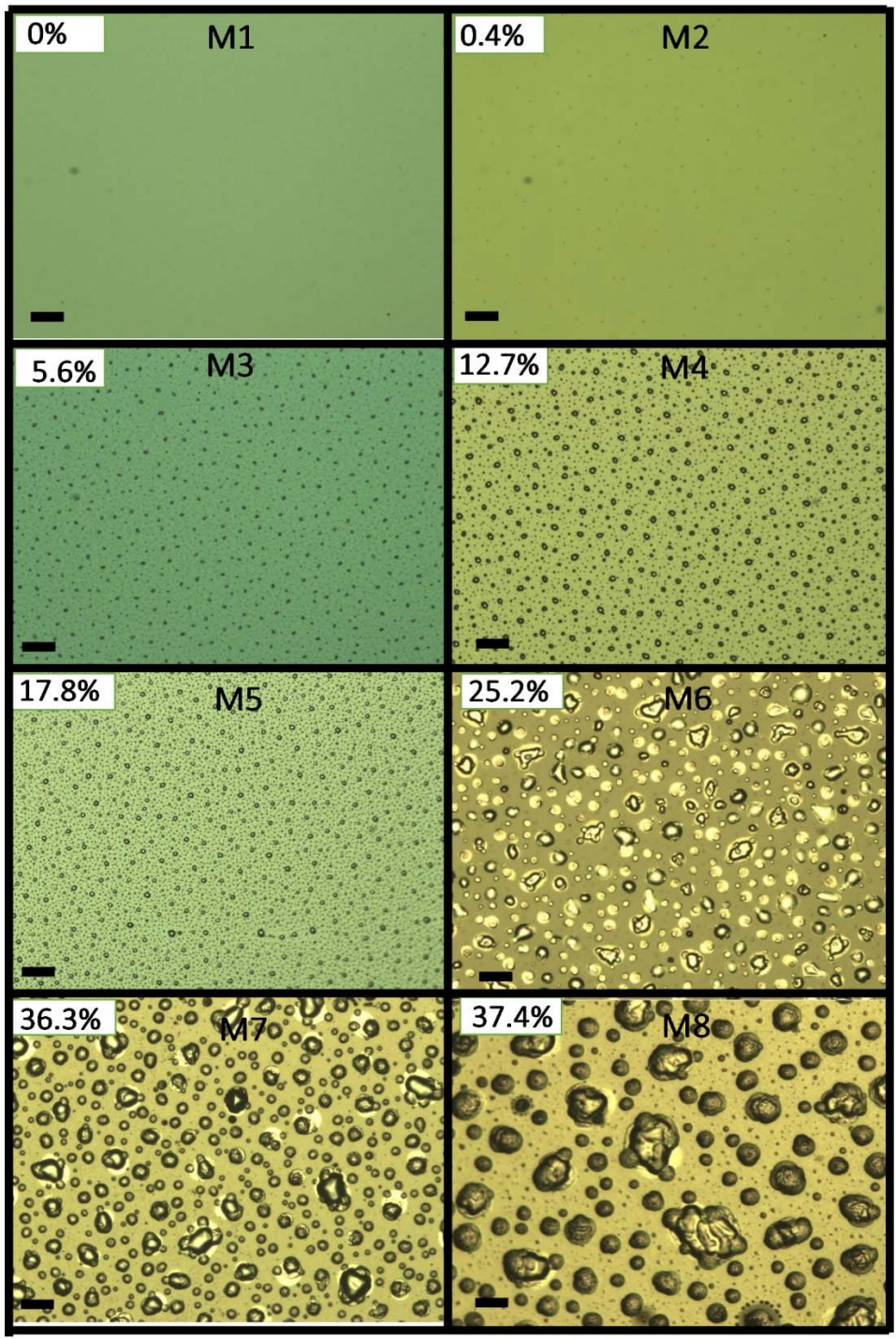

Fig. 2. Optical images of sample surfaces with the indication of the actual surface coverage by droplets. Black line length in all the images is 100 μm.

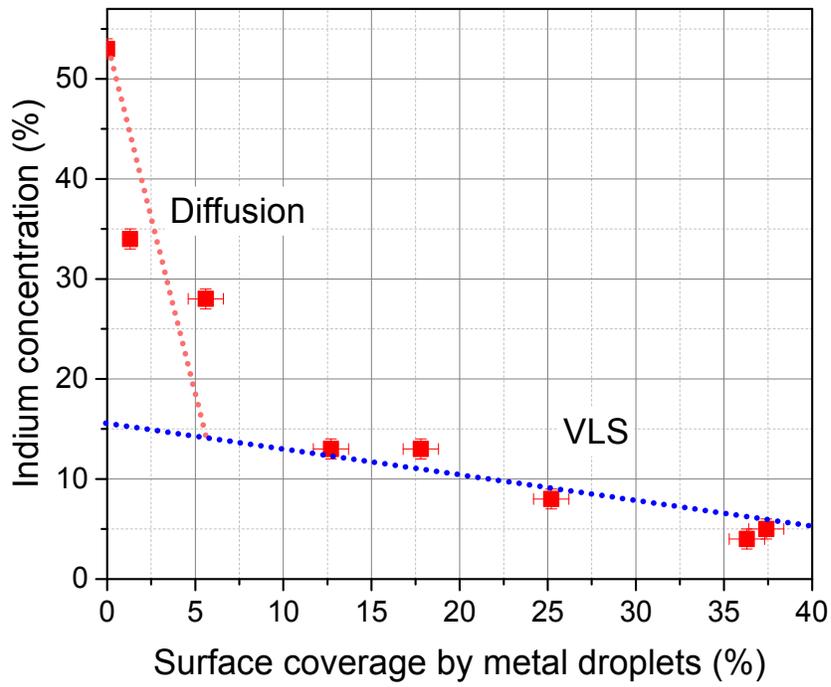

Fig. 3. Indium concentration in InGaN layers (measured by XRD) as a function of metal droplets surface coverage

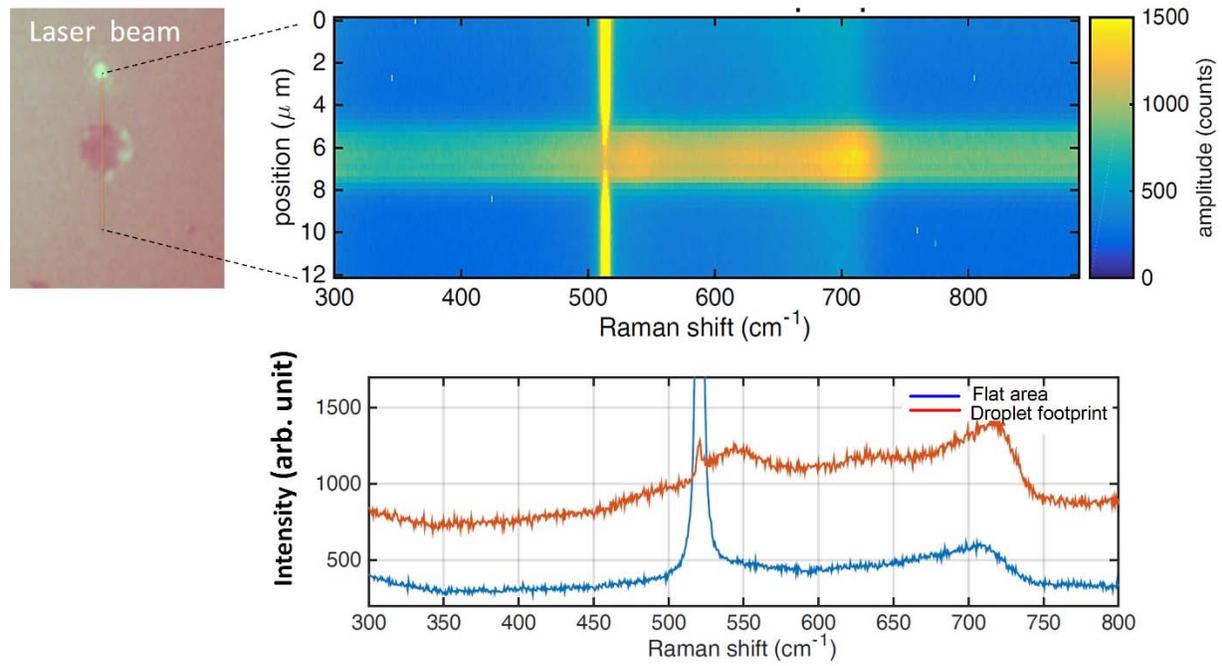

Fig. 4. Line scan Raman spectroscopy around footprint of a metal droplet

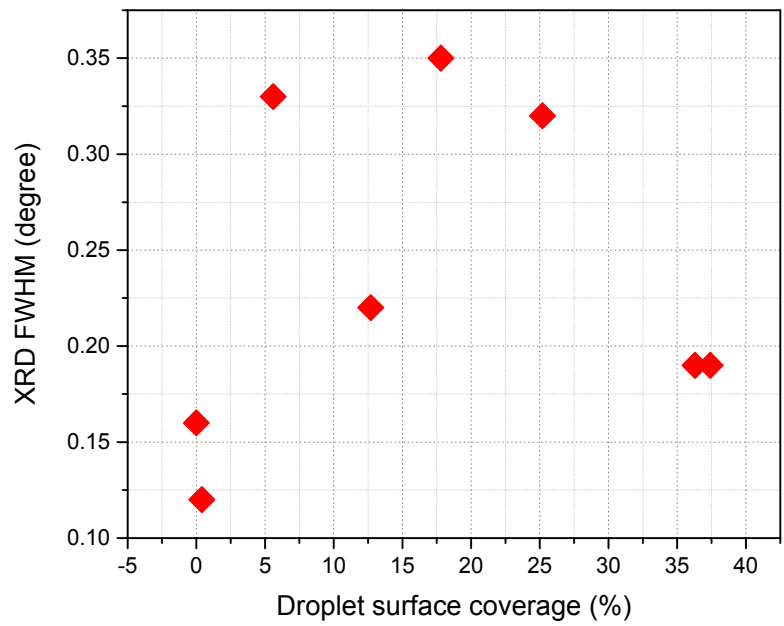

Fig. 5. FWHM of XRD ω-2θ scan of InGaN samples vs surface coverage by metal droplets